\documentclass{PoS}

\title{Few-Body Systems and the Pionless Effective Field Theory}

\ShortTitle{The Pionless EFT}

\author{\speaker{Lucas Platter}\thanks{previous address: Department of
  Physics, Ohio State University, Columbus, OH 43210, USA}\\
        Institute for Nuclear Theory\\
University of Washington\\
Seattle, WA 98195, USA\\
        E-mail: \email{lplatter@phys.washington.edu}}

      \abstract{The pionless effective field theory (EFT) is the
        appropriate low-energy EFT for short-range interactions that
        display a large scattering length. It has been successfully
        applied in atomic, nuclear and particle physics. We give an
        overview over recent calculations employing the pionless
        effective field theory and lay emphasis on applications in the
        three- and four-body sector where the most exciting
        developments have occurred.}

\FullConference{6th International Workshop on Chiral Dynamics, CD09\\
		July 6-10, 2009\\
		Bern, Switzerland}

\begin{document}

\section{Introduction}
As the lightest exchange particle of the internuclear interaction the
pion plays a central role in the conventional description of the
nucleon-nucleon ($NN$) interaction.  However, at sufficiently low
energies the pionic degrees are irrelevant to the description of the
interaction and the $NN$ interaction appears pointlike. In this regime
the effective range parameters can be used to achieve an excellent
description of two-nucleon scattering data. In the case of nucleons
but also in many atomic systems, the effective range parameters
display furthermore a separation of scales between the large two-body
scattering length $a$ and the range $R$ of the interaction which
provides an excellent starting point for the construction of an
effective field theory (EFT) in which quantities are expanded in the
small parameters $a/R$ and $k R$ ($k$ denoting the momentum scale of
the process under consideration). The resulting EFT which in nuclear
physics is known as the pionless EFT, is a framework which facilitates
the model-independent calculation of low-energy observables in systems
with short-range interactions and a large scattering length. In the
two-nucleon system it has been used successfully to calculate
electroweak observables.

In the three-body sector this EFT has provided a new perspective on findings
made in the 1970s. At this time Vitaly Efimov discovered that the zero-range
limit of the 3-body problem for nonrelativistic particles with short-range
interactions shows discrete scale invariance.  If $a = \pm \infty$, there are
infinitely many 3-body bound states with an accumulation point at the 3-atom
scattering threshold. These {\it Efimov states} or {\it Efimov trimers} have a
geometric spectrum \cite{Efimov70}.  Furthermore, he pointed out that these
results were also valid for finite scattering length as long as $a\gg
R$. The EFT analysis has showed that this discrete scale invariance is
associated with a particular renormalization group behavior of the three-body
problem (limit cycle). The phenomena associated with the implications of these
results are generally known as Efimov physics~\cite{Braaten:2004rn} (see
Ref.~\cite{Platter:2009gz} for a summary of recent developments).

Another promising arena for the pionless EFT are halo nuclei which are nuclei
consisting of a tightly bound halo and a small number of additional nucleons
that are weakly bound to the core and form the {\it halo}. One characteristic
of a halo nucleus is that the radius of the halo nucleus significantly larger
than the radius of the core of the halo. This indicates a large scattering
length and therefore a fine-tuned interaction between core and halo nucleons.
The pionless EFT can offer in this context a new perspective on conventional
cluster models which have been used frequently to describe such systems.

In the following section we will outline briefly the key ingredients,
Lagrangian and powercounting, of the pionless EFT. In section \ref{sec:3body},
we will summarize the results when this EFT is applied to the three- and
four-body system and in section \ref{sec:nlo} we report on current efforts to
understand the impact of finite range corrections on predictions for few-body
observables. In section \ref{sec:halo} we will discuss briefly recent results
obtained for halo systems and we will end with a short summary.
\section{The Pionless EFT}
The pionless EFT is the appropriate low-energy theory for reactions
between particles interacting through short-range interactions of
range $R$ at momenta with $k R\ll 1$. The most general Lagrangian describing
such systems is given by
\begin{eqnarray}
  \label{eq:Lagrange1}
\mathcal{L}&=&{\psi}^{\dagger}\left[i
  \partial_t+\frac{{\nabla}^2}{2m}\right]\psi
-\frac{C_0}{4}(\psi^\dagger\psi)^2
-\frac{D_0}{36}(\psi
^{\dagger}\psi)^3-\frac{E_0}{576}(\psi
^{\dagger}\psi)^4...,
\end{eqnarray}
where the ellipses represent operators of higher dimension which means terms
with more derivatives and/or more fields. Here, we have neglected relativistic
effects which are suppressed by factors of $(p/M)^2$.  $D_0$ and $E_0$ denote
the leading three- and four-body interactions.

Depending on the relative size of the effective range parameters,
different powercountings have to be employed for the calculation of
observables. For example if the scattering length is of natural size
($a \sim R$) the powercounting is completely perturbative and only a
finite number a number of diagrams has to be evaluated at every order
in the EFT expansion. However, if the scattering length is large
compared to the range of the interaction ($a \gg R$) the problem
becomes nonperturbative and all connected diagrams that include only
the $C_0$ vertex have to be summed up at leading order in the EFT
expansion. Here, we will focus on the latter case which is of more
interest in the few-body sector. An overview over calculations in the two-body
sector can be found in Ref.~\cite{Bedaque:2002mn}.

\section{Few-Body Systems}
\label{sec:3body}
\subsection{The Three-Body System}
It was shown by Bedaque, Hammer and van Kolck how the pionless EFT is
applied to a three-body system of identical bosons
\cite{Bedaque:1998kg,Bedaque:1998km}. Using an auxiliary field $T$,
they rewrote the Lagrangian given in Eq.~(\ref{eq:Lagrange1})
\begin{equation}
\label{eq:Lagrange2}
\mathcal{L}={\psi}^{\dagger}\left(i
  \partial_t+\frac{\overrightarrow{\nabla}^2}{2m}\right)\psi
+\Delta T^\dagger T-\frac{g}{\sqrt{2}}(T^\dagger\psi\psi+\hbox{h.c.})
+h T^\dagger T\psi^\dagger\psi \ldots .
\label{eq:lagrangian2}
\end{equation}
The Lagrange density above is equivalent to the density in
Eq.~(\ref{eq:Lagrange1}) if the low-energy constants are chosen to be $2
g^2/\Delta=C_0$ and $-18 hg^2/\Delta^2=D_0$ (and the four-body force terms has
been omitted). The advantage of using this formulation is that it turns the
three-body problem in an effective two-body problem. Feynman rules derived
from Eq.~(\ref{eq:Lagrange2}) can be used to derive an integral equation for
particle-dimer scattering. After $S$-wave projection and multiplication with
wave function renormalization factors, the fully-off-shell equation takes the
form
\begin{eqnarray}
  \label{eq:3bdytmatrixswave}
\nonumber
  t(p,k;E)&=&\frac{8\pi}{m a}\biggl[\frac{1}{p k}\ln\left(\frac{p^2+p k+k^2-m E}{p^2-p
      k+k^2-m E}\right)+\frac{2 H(\Lambda)}{\Lambda^2}\biggr]\\
&&\hspace{-1.5cm}+\frac{2}{\pi}\int_0^{\Lambda}\hbox{d}q\, q^2
\frac{t(q,k;E)}{-1/a+\sqrt{3 q^2/4-m E-i\epsilon}}
\left[\frac{1}{p q}\ln\left(
\frac{p^2+p q+q^2-mE}{p^2-p q+q^2-mE}\right)
+\frac{2 H(\Lambda)}{\Lambda^2}\right]~.
\end{eqnarray}
Here a cutoff $\Lambda$ has been introduced to make the integral
equation well-defined and $h=2 m g^2 H(\Lambda)/\Lambda^2$.
Equation~(\ref{eq:3bdytmatrixswave}) is then related to the atom-dimer
phase shift via
\begin{equation}
  t_0(k,k)=\frac{3\pi}{m}\frac{1}{k\cot\delta_{AD}-ik}~.
\end{equation}
Equation (\ref{eq:3bdytmatrixswave}) (without the three-body force) is
also known as the Skorniakov-Ter-Martirosian (STM) equation, named
after the first ones to derive an integral equation for the three-body
problem with zero-range two-body interactions~\cite{STM57}. The
three-body force has to be included however, since without it
observables display strong cutoff dependence. The resulting running of
the three-body force with the cutoff $\Lambda$ shows limit cycle
behavior \cite{Bedaque:1998kg,Bedaque:1998km}.

\subsection{The Four-Body System}
Since the three-body system requires an additional three-body datum
for renormalization it is natural to ask whether the same will happen
in the four-body system and a new four-body observable will be
required for consistent renormalization. This question was answered in
Ref.~\cite{Platter:2004qn}, where an analysis of four-body observables showed that
one two-body and one three-body input are sufficient to obtain
model-independent predictions for four-body observables. This implies
that for fixed scattering length the binding energy of a four-body
bound state will only depend on the value of the three-body observable
used for the renormalization of the three-body system. The pionless
EFT explains therefore the well-known correlation between triton and
$\alpha$-particle binding energy also known as the Tjon line.

This approach was used furthermore for a more detailed analysis of the
four-boson system with large positive and large negative scattering
length~\cite{Hammer:2006ct}.  Results in this analysis also lead to the
conclusion that every trimer state is tied to two universal tetramer states
with binding energies related to the binding energy of the next shallower
trimer:
\begin{equation}
  \label{eq:4body1}
  E_{4,0}\sim 5\, E_T\quad {\rm and} \quad  E_{4,1}\sim 1.01\, E_T
\quad{\rm for}\quad \gamma\sim 0~,
\end{equation}
where $E_{4,0}$ denotes the binding energy of the deeper of the two tetramer
states and $E_{4,1}$ the shallower of the two.

A recent calculation by von Stecher, d'Incao and Greene \cite{Stecher:2008}
supports the findings made in \cite{Platter:2004qn,Hammer:2006ct}. The authors
of this work extended previous results to higher numerical accuracy. They
furthermore considered the relation between universal three- and four-body
bound states in the exact unitary limit ($a\rightarrow\infty$). They found
 \begin{equation}
  \label{eq:4body2}
  E_{4,0}\approx 4.57 E_T\quad {\rm and} \quad  E_{4,1}\approx 1.01 E_T~,
\end{equation}
which agree with the results obtained in Ref.~\cite {Hammer:2006ct} and given
in Eq.~(\ref{eq:4body1}).

\begin{figure}[tb]
\centerline{\includegraphics*[width=10cm,angle=0]{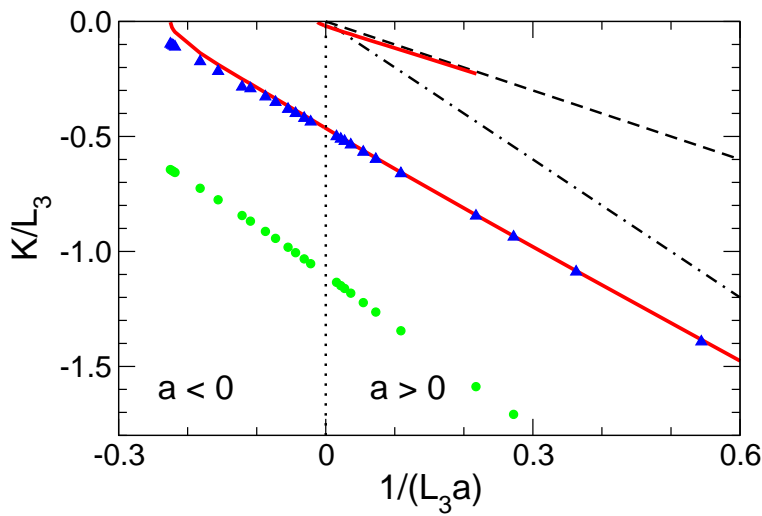}}
\caption{\label{fig:efimov-4body} The $a^{-1}$--$K$ plane for the four-body
  problem. The circles and triangles indicate the four-body ground and excited
  state energies $B_4^{(0)}$ and $B_4^{(1)}$, while the lower (upper) solid
  lines give the thresholds for decay into a ground state (excited state)
  trimer and a particle. The dash-dotted (dashed) lines give the thresholds
  for decay into two dimers (a dimer and two particles).  The vertical dotted
  line indicates infinite scattering length.  All quantities are given in
  units of the three-body parameter $L_3$.  }
\end{figure}

The results obtained by Hammer and Platter in
Ref.~\cite{Hammer:2006ct} were furthermore presented in the form of an
extended Efimov plot, shown in Fig.~\ref{fig:efimov-4body}. Four-body
states have to have a binding energy larger than the one of the
deepest trimer state. The corresponding threshold is denoted by lower
solid line in Fig.~~\ref{fig:efimov-4body}.  The threshold for decay
into the shallowest trimer state and an atom is indicated by the upper
solid line. At positive scattering length, there are also scattering
thresholds for scattering of two dimers and scattering of a dimer and
two particles indicated by the dash-dotted and dashed lines,
respectively.  The vertical dotted line denotes infinite scattering
length. A similar but extended version of this four-body Efimov plot
was also presented by Stecher, d'Incao and Greene in
Ref.~\cite{Stecher:2008}.
They computed also the scattering lengths at which the binding energies of
the tetramer states become zero and found
\begin{equation}
\label{eq:tetra-scatteringlengths}
  a^*_{4,0}\approx 0.43 a_*\quad{\rm and}\quad a^*_{4,1}\approx 0.92 a_*~.
\end{equation}
The authors concluded that at these values of the two-body scattering
length the existence of the universal tetramer states should become
visible as loss features due to recombination processes in systems of
ultracold atoms.

Ferlaino {\it et al.} recently studied the four-body problem with
short-range interactions experimentally~\cite{Ferlaino:2009}. Using
ultracold $^{133}$Cs atoms in the lowest hyperfine state at a
temperature of 50~nK, they found loss features at scattering lengths
$-730 a_0$ and $-410 a_0$ which were interpreted as the four-body loss
features predicted by Stecher, d'Incao and Greene \cite{Stecher:2008}.
With the triatomic Efimov resonance measured at $-870 a_0$, this gives
for the ratios of the four- and three-body resonance position
\begin{equation}
    a^*_{4,0}/ a_* \approx 0.47\quad{\rm and}\quad a^*_{4,1}/ a_*\approx 0.84~.
\end{equation}
These experimental results are in fact surprisingly close to the zero-range
prediction made in \cite{Stecher:2008} since finite range effects are expected
to be important at these values of the scattering length. The range of the Cs-Cs
interaction (which is set by the van-der Waals length scale) is approximately
$200 a_0$.

\section{Higher Order Corrections}
\label{sec:nlo}
The promise of EFTs is that observables can be calculated to high
accuracy. To deliver that promise is has first to be understand which
operators have to be taken into account at what order. While the
required two-body operators follow trivially from the effective range
expansion, it is not a priory clear at what order the next three-body
force enters. Hammer and Mehen calculated the phaseshift of
neutron-deuteron scattering up to next-to-leading order (NLO)
perturbatively and demonstrated that no additional three-body
parameter is required \cite{Hammer:2001gh} (as long as the scattering
remains fixed as we will discuss below). An analysis of the necessity
of three-body forces in higher partial waves was carried out by
Grie\ss hammer \cite{Griesshammer:2005ga}.  Different conclusion have
been reached for next-to-next-to-leading (N2LO) order. In
Ref.~\cite{Bedaque:2002yg} it was found that a new energy-dependent
three-body force is required. A renormalization group analysis of the
large cutoff behavior of the three-body amplitude lead the authors of
Ref.~\cite{Platter:2006ev} to the conclusion that an energy-dependent
counterterm would be required at N3LO. In both references, the kernel
of the three-body integration was modified to account for the effects
of the effective range. The reason for the disagreement between both
results might therefore simply be the fact that the cutoff dependence
of the three-body amplitude is different for large cutoffs than for
natural cutoffs ($\Lambda \sim 1/R$).

In a recent work \cite{Platter:2008cx} the next-to-leading order
correction was reconsidered using a perturbative approach. The fact
that the three-body bound state wave function is known exactly for
infinite scattering length was used to calculate the NLO shift exactly
in this limit. It was shown that the bound state spectrum receives in
this case no correction and that the discrete scale invariance of the
three-body wave function has therefore a direct effect on the size of
higher order corrections. A future publication will discuss the need
of an additional energy-independent three-body counterterm appearing
at NLO that is proportional to the inverse scattering length
\cite{Ji:2009}. This counterterm will only be relevant for the NLO
analysis of problems where the scattering length is variable such as
in experiments that measure the three-body recombination rate around a
Feshbach resonance.

The impact of range corrections in the four-body sector was discussed
by Kirschner {\it et al.} in Ref.~\cite{Kirscher:2009aj}.

\section{Halo Nuclei}
\label{sec:halo}
Halo nuclei are another possible application of the EFT for
short-range interactions. The weak binding of the halo nucleons to the
core nucleus indicates a separation of scales which might be
understood in terms of a large core-nucleon scattering length. The
first application of the short-range EFT to halo nuclei was carried
out in Refs.~\cite{Bertulani:2002sz,Bedaque:2003wa}. In these works
the authors considered the one-neutron halo $^5$He and calculated
phaseshifts and cross sections for elastic $\alpha$-nucleon
scattering. A further example of a nuclear two-body cluster that has
been considered is the 2-$\alpha$ system~\cite{Higa:2008dn}.

\begin{figure}[t]
\centerline{\includegraphics*[width=4.in,angle=0]{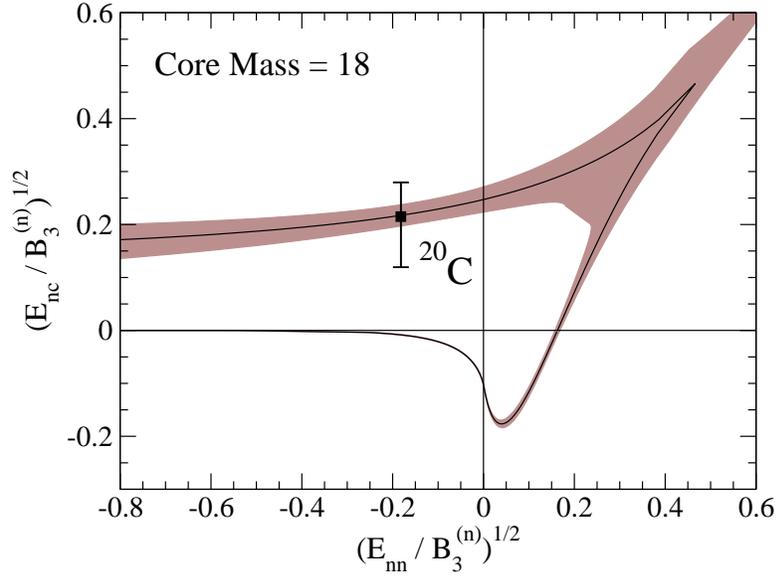}}
\caption{\label{fig:halo} Boundary curve in the $\sqrt{E_{nc}/B_3^{(n)}}$
  vs. $\sqrt{E_{nn}/B_3^{(n)}}$ plane with leading order error bands.
  Boundary curve shown for a core mass of $A = 18$ with the experimental data
  for $^{20}$C from Ref.~\cite{TUNL}. Figure taken from Ref.~\cite{Canham:2008jd}}
\end{figure}
Recently, Canham and Hammer~\cite{Canham:2008jd} performed the first
EFT calculation for two-neutron halos, i.e. the three-body case. In
their work they calculated the binding energies and radii of halos
such $^{11}$Li and $^{20}$C. Canham and Hammer also addressed the
question whether any of the considered systems supports an excited
Efimov state.  Fig.~\ref{fig:halo} shows a parametric plot
($(E_{nc}/B_3^{(n)})^{1/2}$ versus $(E_{nn}/B_3^{(n)})^{1/2}$) which
describes the region in the two-body parameter space that supports a
three-body state above $B_3^{(n)}$. They found that the $^{20}$C
system might exhibit an excited Efimov state close to the threshold.

\section{Summary}
We have discussed recent applications of the pionless EFT to few-body
systems. Predictions for few-body observables can be made with this
EFT provided one three-body datum is know. However, even in the
absence of such an input, the pionless EFT is capable of explaining
correlations between few-body observables (e.g, the Tjon line). Its
success demonstrates that it is the ideal tool to analyze the
universal properties of systems with a large scattering whether in the
atomic or nuclear context.

It is furthermore a small parameter expansion that
promises high accuracy for electroweak observables of wide
interest. Form factors \cite{Platter:2005sj,Sadeghi:2009dm}, thermal
neutron capture on the deuteron \cite{Sadeghi:2005aa,Sadeghi:2006aa}
and triton photo-dissociation \cite {Sadeghi:2009rf} have already been
considered. However, other observables such as triton $\beta$-decay
or electroweak reactions in the four-body sector remain to be calculated.

Halo nuclei are a relatively new application of the pionless EFT. Here
it can provide answers to questions such as whether Halo nuclei are
examples of Efimov physics and whether these states might display
additional excited states belonging to an Efimov spectrum. Since the
EFT for short-range interactions also significantly simplifies the
complexity of this problem (for a example in the case of $^{20}$C
a 20-particle problem is reduced to a 3-body problem) one can hope that
the EFT treatment of halo systems will also facilitate a calculation
of scattering observables.

\section*{Acknowledgments}
I thank the organizers of the conference for a very interesting
workshop and providing a stimulating environment. This work was
supported in part by the National Science Foundation under Grant
No.~PHY--0653312, the UNEDF SciDAC Collaboration under DOE Grant
DE-FC02-07ER41457 and the Department of Energy under grant number
DE-FG02-00ER41132.

\end{document}